# Supramolecular Co(II)-[2 × 2] grids: Metamagnetic behavior in a single molecule

*Oliver Waldmann,*[*,†] *Mario Ruben,*[‡] *Ulrich Ziener,*[§] *Paul Müller,*[†] *Jean M. Lehn*[∥]

[†]Physikalisches Institut III, Universität Erlangen-Nürnberg, 91058 Erlangen, Germany.

[‡]Institut für Nanotechnologie, Forschungszentrum Karlsruhe, 76021 Karlsruhe, Germany.

[§]Organische Chemie III/Makromolekulare Chemie, Universität Ulm, 89069 Ulm, Germany.

[∥]ISIS, 8 Alleé Gaspard Monge, 67083 Strasbourg, France.

Dr. Oliver Waldmann, Department of Chemistry and Biochemistry, University of Bern, Freiestrasse 3, CH-3012 Bern, Switzerland. Email: waldmann@iac.unibe.ch

The magnetic anisotropy of the supramolecular [2 × 2] grid $[Co(II)_4L_4]^{8+}$, with a bis(bipyridyl)-pyrimidine based ligand L, was investigated by single-crystal magnetization measurements at low temperatures. The magnetization curves exhibit metamagnetic like behavior and are explained by the weak-exchange limit of a minimal spin Hamiltonian including Heisenberg exchange, easy-axis ligand fields, and the Zeeman term. It is also shown that the magnetic coupling strength can be varied by the substituent $R_1$ in the 2-position on the central pyrimidine group of the ligand L.

## 1. Introduction

The design of small magnetic clusters has become a major goal in the area of nanoscale materials as these objects promise novel magnetic properties. For instance, quantum tunneling of the magnetization has been observed in molecular nanomagnets like $Mn_{12}$ or $Fe_8$.[1] Supramolecular chemistry provides a unique tool to produce, by self-assembly, molecular architectures with a defined geometry of metal centers.[2] The so-called [N × N] grids attracted particular interest: the arrangement of exactly $N^2$ metal centers in a flat, regular N × N matrix suggests applications in e.g. information storage and processing technology.[3,4] Numerous magnetic [2 × 2] and [3 × 3] grids with Fe(II), Co(II), Ni(II), or Cu(II) ions were created, and magnetic studies demonstrated a remarkable variety of their magnetic properties.[5,6]

In a previous study we demonstrated that Co(II)-[2 × 2] grids based on bis(bipyridyl)-pyrimidine ligands (see Figure 1) exhibit intramolecular antiferromagnetic (AF) interactions on the order of Kelvins, while intermolecular magnetic interactions are negligibly small (estimated to be at best ≈ 10 mK).[5] The magnetism of macroscopic samples thus reflects the magnetic properties of single molecules. In this work, the [2 × 2] grid molecule **1** (Figure 1) is explored by means of single-crystal magnetization measurements unraveling a metamagnetic-like behavior linked to an effective Ising-type interaction of the metal centers. In addition, from the magnetism of powder samples of the Co(II)-[2 × 2] grids **2** and **3** (Figure 1) it is demonstrated that the AF coupling constant depends on the substituent $R_1$ on the ligand L, i.e, can be tuned by a controlled chemical variation of the ligands. For the understanding of the magnetism in the Co(II)-[2 × 2] grids it is useful to first analyze carefully the magnetism in the mononuclear analog $[Co(II)(tpy)_2](PF_6)_2$ (**4**).

## 2. Experimental Methods

The magnetic properties of powder or single-crystal samples were measured with a commercial SQUID magnetometer (Quantum Design). The details of the measurement procedures were as described in Ref. 6a. The weight of the powder samples was typically 1-3 mg. The weight of the single-crystal samples was only ≈ 50 μg; the magnetic measurements were thus difficult. In particular, because of their smallness, the accuracy of the orientation of the crystal samples with respect to the magnetic field was modest (≈ 15°).

## 3. Results

**3.1. Syntheses and Description of Some Structural Details.** The grid complexes **1**, **2**, **3**, and the mononuclear analogue [Co(II)(tpy)$_2$](PF$_6$)$_2$ (**4**) were synthesized following known literature procedures.[3,7] Also samples were prepared, where the positive charges of the grid clusters were countered by BF$_4^-$ ions; the magnetic properties were found to be independent on the counter ions.

The structure of the [2 × 2] grids [Co(II)$_4$L$_4$]$^{8+}$ consists of the four bis(bipyridyl)-pyrimidine based ligands L and four Co(II) metal centers, see Figure 1. Each metal center is situated at the crossing point of two ligands and is enclosed by six N atoms in an octahedral geometry. The distance of separation of the Co(II) centers is about 6.5 Å. The molecules exhibit an approximate D$_2$ symmetry. The steric requirements of the ligand induce a pronounced tetragonal compression of the coordination sphere formed by the N atoms surrounding a Co(II) center in a direction perpendicular to the grid plane, and a slight orthorombic distortion. For grid **1**, the average Co-N bond lengths are 2.16 Å (N of the terminal pyridine), 2.23 Å (N of the pyrimidine), and 2.03 Å (remaining N).[3] This is of relevance for the understanding of the magnetism (*vide infra*).

**3.2. Magnetic Measurements.** The temperature-dependent magnetic susceptibilities and the low-temperature magnetization curves of powder samples of **1**, **2**, and **3** are shown in Figure 2 (for **1** similar curves were presented already in Ref. 5). The general magnetic behavior of the three grids is apparently very similar to each other. In the following, we thus concentrate the discussion on **1**. The analysis of the situation in **2** and **3**, to be given in section 4.4, follows the same logic of arguments and is then straightforward.

The magnetic susceptibility of **1** exhibits a maximum at $T^* \approx 7.5$ K due to an intramolecular AF coupling.[5] The magnetization curve shows an unusual behavior: it increases at first almost linearly with the magnetic field, rises then to an inflection point at $B^* \approx 3.5$ T, and starts to saturate at higher fields. This behavior is reminiscent of a thermally broadened magnetization step at $B^*$ due to a field-induced ground state level-crossing, as it is often observed in AF clusters.[8] If this were the whole story, however, an exponential instead of a linear behavior would be expected at fields below the level-crossing field $B^*$. This point is of importance in the following.

For **1**, also single-crystal measurements were performed. Figure 3a presents the field dependence of the magnetic moment of a single crystal of **1** at 1.9 K for three different orientations of the magnetic field. A pronounced magnetic anisotropy is apparent. The magnetic moment for fields $B \parallel z$, $m_z$, exhibits a thermally broadened, but clear magnetization step at $B^* = 3.5$ T. At small fields, $m_z$ increases exponentially with field as expected for a magnetization step. In contrast, the magnetic moments for fields in the x and y directions, $m_x$ and $m_y$, increase linearly with equal slope up to a field of about 3 T. The deviation of $m_x$ and $m_y$ from linear behavior above 3 T is reminiscent of a small contribution of $m_z$ in these measurements, which we explain by a small misalignment of the crystal with respect to the magnetic field (which is smaller for $m_x$ than for $m_y$). The $m_x$ and $m_y$ data is thus interpreted as to show i) a linear increase of the magnetization with field and ii) that the magnetic anisotropy in the xy direction is negligible (justifying a model with tetragonal symmetry for the analysis of the data).

## 4. Analysis and Discussion

Our previous study demonstrated a high-spin state of the Co(II) centers in the grid **1** and in **4**.[5] The magnetic properties of Co(II) ions are notoriously difficult to describe because of the orbital contribution. It is hence useful to analyze at first the magnetism in the mononuclear analog **4**.

**4.1. Magnetism in [Co(II)(tpy)$_2$](PF$_6$)$_2$ (4).** In the crystal-field and LS-coupling approximation, the Hamiltonian of the lowest LS multiplet of a single Co(II) center is $\hat{H}_{LS} = V + K + \lambda \hat{L} \cdot \hat{S} + \mu_B(\hat{L} + 2\hat{S}) \cdot \mathbf{B}$.[9] The crystal field has been decomposed into an octahedral part V and fields of lower symmetry K. $\lambda$ denotes the spin-orbit coupling constant, $\mu_B$ the Bohr magneton, and **B** the magnetic field vector. The dominant octahedral ligand field V leads to a $^4T_1$ ground-state multiplet. Its splitting due to low-symmetry ligand fields, spin-orbit coupling, and magnetic field is generally treated by first-order perturbation theory, which is equivalent to replacing the orbital angular-momentum operator $\hat{L}$ by a pseudo angular-momentum operator $\hat{l}$, with $\hat{L} = -3/2\hat{l}$ and l = 1.[10] The effective Hamiltonian for tetragonal symmetry then becomes

$$\hat{H}_1 = \delta \hat{l}_z^2 - \frac{3}{2}\alpha\lambda \hat{l}_z \hat{S}_z - \frac{3}{2}\alpha'\lambda(\hat{l}_x \hat{S}_x + \hat{l}_y \hat{S}_y) + \mu_B(-\frac{3}{2}\beta \hat{l}_x + 2\hat{S}_x)B_x + \mu_B(-\frac{3}{2}\beta'\hat{l}_z + 2\hat{S}_z)B_z \qquad (1)$$

Here, we have introduced $\alpha$ and $\beta$ factors to account for i) the mixing of the $^4T_1(^4F)$ ground state with the next higher lying $^4T_2$ level,[10] ii) the mixing with the $^4T_1(^4P)$ excited state,[11] iii) an orbital reduction factor due to covalent bonding effects, and iv) reduction of the spin-orbit coupling, also due to covalency.[12] The parameters were chosen such that $\alpha, \alpha', \beta, \beta' \leq 1$ (with reduction ii only the parameters assume 1 in the weak-field limit and 2/3 in the strong-field limit). Furthermore, $\alpha \leq \beta$ and $\alpha' \leq \beta'$ since reduction iv is not relevant for the Zeeman terms.

This approach implicitly assumes a modest strength of the low-symmetry fields with respect to the spin-orbit coupling, i.e., $K \approx \lambda$. However, the pronounced tetragonal compression of the coordination spheres in **4** and the grids **1-3** in combination with the small free spin-orbit coupling of Co(II) ($\lambda$ = 172 cm$^{-1}$) suggests the limit $K \gg \lambda$. This is also suggested by the exceptional strong g-factor anisotropy of the lowest Kramers doublet found in **4** ($g'_{xy}$ = 1.25, $g'_z$ = 7.0).[5] In this limit, the orbital contribution could be quenched, implying the usual effective spin Hamiltonian for orbitally non-degenerate ions,[9]

$$\hat{H}_2 = D\hat{S}_z^2 + E(\hat{S}_x^2 - \hat{S}_y^2) + \mu_B g_{xy}(\hat{S}_x B_x + \hat{S}_y B_y) + \mu_B g_z \hat{S}_z B_z. \tag{2}$$

We least-square fit the data obtained for powder samples of **4** in the temperature range 2-300 K at fields of 0.5, 2, 4, and 5.5 T to both $\hat{H}_1$ and $\hat{H}_2$ (not shown). $\hat{H}_2$ produced very good fits to the whole data set (D = -85 K, E = -20 K, $g_{xy}$ = 2.2, $g_z$ = 2.35). $\hat{H}_1$ led to similar good fits, but for several parameter sets. For $\delta > 0$, we found $\delta$ = 204 K, $\alpha$ = 0.22, $\alpha'$ = 0.802, $\beta$ = 0.514, $\beta'$ = 0.622. For $\delta < 0$, several sets were found, but only $\delta$ = -69.7 K, $\alpha$ = 0.212, $\alpha'$ = 0.444, $\beta$ = 0.271, $\beta'$ = 0.565 did not violate one of the restrictions for the $\alpha$ and $\beta$ factors, though their values are unusually small. It is often observed for Co(II) complexes that both magnitude and sign of $\delta$ are not determined unambiguously.[11] As a summary, both $\hat{H}_1$ and $\hat{H}_2$ describe the data well and a final distinction is difficult, though by stability and reasonability of parameters one might favor $\hat{H}_2$.

The indifference of $\hat{H}_1$ and $\hat{H}_2$ is explained as follows. $\hat{H}_1$ produces six Kramers doublets which are separated in energy by the spin-orbit coupling and ligand field; $\hat{H}_2$ leads to two Kramers doublets with a gap of $2\sqrt{D^2 + E^2}$. It is common practice to introduce an effective spin $S' = 1/2$ and a corresponding effective spin Hamiltonian $\hat{H}' = \mu_B g'_{xy}(\hat{S}'_x B_x + \hat{S}'_y B_y) + \mu_B g'_z \hat{S}'_z B_z$ to describe the lowest doublet (again assuming tetragonal symmetry).[9,10] With appropriate parameters, both $\hat{H}_1$ and $\hat{H}_2$ produce exactly the same effective Hamiltonian $\hat{H}'$ for the ground state doublet - and thus also exactly the same low-temperature properties. So, $\hat{H}_1$ and $\hat{H}_2$ differ only in the higher lying spectrum, but this is discriminated only poorly by magnetization measurements at higher temperatures, because then a thermal average of essentially all states is measured. In some sense, $\hat{H}_2$ is just an effective Hamiltonian for the two lowest Kramers doublets only. We will use $\hat{H}_2$ to describe the single-ion properties but emphasize, that this is more a matter of convenience than a physical statement.

**4.2. Magnetism in the Co(II)-[2 × 2] Grid 1.** Having clarified how to treat a single Co(II) ion, we proceed with analyzing the magnetism of the Co(II)-[2 × 2] grid **1**. As with the single-ion properties, the description of intramolecular magnetic interactions is complicated by the orbital contribution of high-spin Co(II) ions. This point is ignored at the moment, and an isotropic Heisenberg exchange interaction is considered. The spin Hamiltonian for Co(II)-[2 × 2] grids then becomes

$$\hat{H}_{2\times 2} = -J\left(\sum_{i=1}^{3}\hat{\mathbf{S}}_i \cdot \hat{\mathbf{S}}_{i+1} + \hat{\mathbf{S}}_4 \cdot \hat{\mathbf{S}}_1\right) + D\sum_{i=1}^{4}\hat{S}_{i,z}^2 + \mu_B g_{xy}(\hat{S}_x B_x + \hat{S}_y B_y) + \mu_B g_z \hat{S}_z B_z \quad (3)$$

where now $\hat{\mathbf{S}} = \sum_{i=1}^{4}\hat{\mathbf{S}}_i$. In view of the approximate $D_2$ molecular symmetry of the Co(II)-[2 × 2] grids, we have set E = 0, in accordance with the experimental finding of $m_x \approx m_y$ in the single-crystal measurements on **1**. The following analysis will show, that the anisotropy is the dominant term while the exchange interaction is a small perturbation.

The observed values for both $T^*$ and $B^*$ in **1** (7.5 K and 3.5 T, respectively) indicate that the coupling constant J is on the order of Kelvins. If the case $|D| \ll |J|$ would be realized, the Heisenberg term would lead to level-crossings at fields $B_n \approx n|J|/(\mu_B g)$,[8] implying J ≈ -2.5 K. But then, steps should not be observable in the magnetization curve, in contrast to experiment, since at 1.9 K temperature and coupling constant are on the same order, $k_B T \approx |J|$, and magnetization steps would be thermally washed out completely. Furthermore, the results for **4** strongly suggest that the zero-field splitting D is on the order of several tens of Kelvins (with D < 0) also in the Co(II)-[2 × 2] grids. Thus, $\hat{H}_{2\times 2}$ should be analyzed in the limit $|J| \ll |D|$.

In this situation, the weak-exchange limit, the Heisenberg term is treated perturbatively (in contrast to the strong-exchange limit, which is realized in virtually all molecular nanomagnets of current interest). In the following, only the low-temperature properties at $k_B T \ll |D|$ are of interest and, again, effective spins $S'_i = 1/2$ are introduced for each center in order to describe the lowest lying single-ion Kramers

doublets. The first-order contributions to the effective spin Hamiltonian were shown to be most easily obtained by the substitution $\hat{S}_{i,\nu} = (g'_\nu / g_\nu)\hat{S}'_{i,\nu}$, where $\nu = x, y, z$.[13] Since D < 0, the low-lying single-ion Kramers doublets consist of the states with the magnetic quantum numbers m = 3/2 and m = -3/2, so that $g'_{xy} = 0$ and $g'_z = 3g_z$. Including second-order contributions (whereby neglecting terms of order J/D), the effective Hamiltonian

$$\hat{H}'_{2\times 2} = -J'\left(\sum_{i=1}^{3}\hat{S}'_{i,z}\hat{S}'_{i+1,z} + \hat{S}'_{4,z}\hat{S}'_{1,z}\right) + \mu_B \sum_{i=1}^{4} g'_z \hat{S}'_{i,z} B_z - \frac{1}{2}\chi'_0 (B_x^2 + B_y^2) \qquad (4)$$

is obtained, with $\chi'_0 = 3\mu_B g_{xy}^2 /(4|D|)$. Importantly, because of $g'_{xy} = 0$, the magnetic interaction is strictly of the Ising type with $J' = (g'_\nu / g_\nu)^2 J = 9J$. Furthermore, the Zeeman term is effective only for fields in the z direction, while the second-order contribution appears only for fields in the xy plane. This leads to a markedly different magnetic behavior for fields parallel and perpendicular to the z axis.

The energy spectrum and magnetic properties of $\hat{H}'_{2\times 2}$ are easily derived. For magnetic fields in the z direction, the energy spectrum is shown in Figure 4b. The Ising interaction leads to a level-crossing at $B_c = |J'|/(\mu_B g'_z)$, at which actually three states cross, namely those with total magnetic quantum number M = 0, -1, -2. One thus expects a pronounced step at the level-crossing field $B_c$ in the magnetization curve, which at finite temperatures is thermally broadened. For fields in the xy plane, however, only the second-order term is active. It results in a magnetic moment, which increases linearly with magnetic field with slope $\chi'_0$. The comparison with experiment is satisfying: All the characteristic features of the above theoretical picture are exactly as found also in the experiment. This is underlined by Figure 3b, which presents a numerical calculation of the magnetic moments using the full Hamiltonian $\hat{H}_{2\times 2}$.

Two points should be noted. At first, it is apparent that the linear increase of the magnetization at lower fields observed in the powder data (see Figure 2b) reflects the linear increase due to the second-order contribution in $\hat{H}'_{2\times 2}$ for fields in the xy plane. This underscores the importance of second-order contributions (which seem to have been neglected generally in analyzes of magnetic data of molecules with orbitally degenerate ions). Secondly, in contrast to the situation of an assumed dominant Heisenberg interaction, the magnetization step is now well resolved even at a measurement temperature of T = 1.9 K because of the "amplification" of the effective coupling constant, $J' = 9J$. Together with $B^* = 3.5$ T, $B_c = |J'|/(\mu_B g'_z)$, and $g'_z = 7.0$ as observed for **4**,[5] the coupling constant is estimated to $J \approx -1.9$ K. The value used for $g'_z$ is consistent with $g_z \approx g = 2.4$ as determined from the room-temperature effective moment $\mu_{eff} = 4.6$ of **1**.[5] The Ising interaction leads to a maximum in the susceptibility at $T_c = 0.6|J'|/k_B$ suggesting $J \approx -1.4$ K, which is in good agreement with the former estimate and demonstrates the consistency of the analysis.

Our analysis, which is based on $\hat{H}_{2\times2}$, does not provide a full quantitative description of the magnetism in the Co(II)-[2 × 2] grid **1**. This is not surprising in view of the many approximations inherent in $\hat{H}_{2\times2}$. The problem concerning the appropriate single-ion spin Hamiltonian has been discussed. Furthermore, the magnetic interactions were described by an isotropic Heisenberg term, an approximation since orbital contributions should not be negligible. However, in any case the low-energy spectrum can be covered by an effective spin Hamiltonian acting in the space spanned by the lowest single-ion Kramers doublets. The most general form allowed by the $D_2$ molecular symmetry for e.g. the magnetic interaction is $J'_{xy}(\hat{S}'_{i,x}\hat{S}'_{j,x} + \hat{S}'_{i,y}\hat{S}'_{j,y}) + J'_z\hat{S}'_{i,z}\hat{S}'_{j,z}$. With regard to the low-temperature properties, the main limitation of $\hat{H}_{2\times2}$ is thus a restriction to pure Ising interactions. A nonzero value of $J'_{xy}$ is conceivable, and in fact suggested by $g'_{xy} = 1.25 \neq 0$ observed for **4**.[5] A comparison of the energy spectra for Ising and Heisenberg interactions (Figure 4) shows, that in the latter case, due to a different zero-field level pattern, two equidistant level-crossings appear. By increasing the anisotropy of the magnetic interaction from Heisenberg to Ising, the two level-crossings approach each other to meet exactly for the Ising case. Having only one level-crossing is thus a fingerprint of pure Ising interactions. Small deviations from $J'_{xy} = 0$ lead to two close level-crossings, which result in an additional non-thermal broadening of the magnetization step at finite temperatures. The good agreement of the experimental and theoretical broadenings in Figure 3 thus indicates, that the pure Ising case is actually very well realized in **1**. We associate this exceptional fact to the peculiar structure of the Co(II) grids: the stiff ligands induce strong distortions of the Co(II) coordination spheres, while the [2 × 2]-grid motif simultaneously enforces a uniaxial magnetic behavior. As a summary, $\hat{H}_{2\times 2}$ should be regarded as the minimal spin Hamiltonian which describes what we consider as the defining characteristics: i) the appearance of one magnetization step for fields in the z direction and ii) a linear magnetization for fields in the xy plane. Both points are linked to the limit $|J| \ll |D|$ with J,D < 0.

**4.3. Interpretation in Terms of Metamagnetic Behavior.** Interestingly, the magnetization curves shown in Figure 3a resemble very closely those of metamagnets.[14] Metamagnetism appears in extended antiferromagnets with an easy-axis magnetic anisotropy exceeding the antiferromagnetic exchange interaction. The general behavior of antiferromagnets can be well understood within a classical picture. A (simple) antiferromagnet consists of two interpenetrating ferromagnetic sublattices, A and B, with opposite magnetizations, $\mathbf{M}_A = -\mathbf{M}_B$, in zero field (the magnitude of the sublattice magnetization vectors is obviously equal, and denoted as $M_0$). The classical magnetic energy is given by

$$E = \mu \mathbf{M}_A \cdot \mathbf{M}_B - K(M_{A,z}^2 + M_{B,z}^2) - (\mathbf{M}_A + \mathbf{M}_B) \cdot \mathbf{B}, \tag{5}$$

where the first term describes the magnetic interaction ($\mu > 0$ for an antiferromagnet), the second term a unixial magnetic anisotropy (K > 0 for easy-axis), and the last term the Zeeman interaction. The analogy with the Hamiltonian $\hat{H}_{2\times 2}$ (eq. 3) is apparent. The classical ground state for each µ, K, and **B** is found from minimization of eq 5. The following situations can be observed (which are schematically presented in Figure 5a): For small magnetic fields along the anisotropy axis z, the sublattice magnetization vectors remain in their antiparallel configuration, and the total magnetization $\mathbf{M} = \mathbf{M}_A + \mathbf{M}_B$ is thus zero. For weakly anisotropic antiferromagnets, characterized by $K < \mu/2$, the system exhibits a spin-flop transition at a field $B_{SF} = 2M_0\sqrt{K(\mu-K)}$, at which the magnetization jumps to a non-zero value $M_z < 2M_0$. The magnetization steadily increases with further increasing field, and reaches saturation, i.e., $M_z = 2M_0$, at $B_{sat} = 2M_0(\mu-K)$. For magnetic fields perpendicular to the anisotropy axis, the sublattice magnetizations continuously tilt towards the magnetic field with increasing field, resulting in a magnetization $M_x = \chi B_x$ with $\chi = 1/(\mu+K)$. This is the behavior typically associated with an antiferromagnet (see Figure 5b). Metamagnetism, in contrast, is observed for systems with large magnetic anisotropy (or small exchange), specifically $K > \mu/2$. In this case, for fields parallel to z, no spin flop occurs, but a spin-flip transition at a field $B_c \equiv \mu M_0$. Here, the system jumps directly from the fully antiparallel to the fully polarized configuration, i.e., from zero magnetization to saturation magnetization $M_z = 2M_0$. For perpendicular fields, the situation is as above and a magnetization $M_x$, which increases linearly with applied magnetic field, is observed. The magnetization curves typical of a metamagnet are sketched in Figure 5c.

The similarity of the metamagnetic behavior with the magnetization curves observed in the Co(II)-[2 × 2] grids is now obvious. In this picture, the magnetization step in $m_z$ corresponds to the spin-flip transition for fields exceeding $B_c$, and the linear increase in $m_x$ and $m_y$ to the tilt of the magnetization vectors by the action of a perpendicular field. In fact, the minimal Hamiltonian $\hat{H}_{2\times 2}$ is just the finite-size version of a spin Hamiltonian frequently used to discuss metamagnetism at the microscopic level.[15] The condition $\mu/2 < K$ corresponds to our finding that $|J| \ll |D|$, because clearly $\mu \propto |J|$ and $K \propto |D|$. Of course, in the Co(II) grids no long-range order develops as intermolecular interactions are negligibly small. Thus, the spin-flip transition at $B_c$ is increasingly washed out with increasing temperature, and does not extent to a finite critical temperature as in "true" metamagnets. However, the magnetism in the Co(II)-[2 × 2] grids is completely describable in the same language as used for the metamagnets (and the above discussion showed that it is actually useful to do so), and in this sense we speak of single-molecule metamagnets.

**4.4. Tuning of Magnetism in the Co(II)-[2 × 2] Grids.** We also investigated the magnetism of powder samples of **2** and **3**, see Figure 2. As noted already, the general behavior of their magnetic properties is very similar to that of **1**, and the above analysis thus applies equally well to them. However, interestingly, both the maximum in the susceptibility and the field position of the magnetization step increase in the series of the grids **1**, **2**, **3**. Since both characteristics are related to the coupling constant J via $T^* \propto |J|$ and $B^* \propto |J|/g$, this is an unambiguous evidence for an increase of the coupling constant by about 50%. The variation is not dramatic, but unequivocally demonstrates the potential of a controlled tuning of the magnetic properties in these systems.

## 5. Conclusions

In conclusion, from single-crystal measurements we demonstrated a metamagnetic-like behavior in the Co(II)-[2 × 2] grid molecules of general formula $[Co(II)_4L_4]^{8+}$, with a bis(bipyridyl)-pyrimidine based ligand L. We analyzed carefully the implications of the orbital degeneracy of high-spin Co(II) ions and established a minimal spin Hamiltonian, which describes the characteristic features very well. Possible improvements were indicated. The Co(II)-[2 × 2] grids thus represent excellent model systems to study the microscopic details involved in metamagnetism, thus enlarging our understanding of this peculiar magnetic effect. Furthermore, the great potential inherent in these systems to tune the magnetic interaction by a controlled chemistry has been demonstrated.

**Acknowledgments**

We thank Dr. D. Bassani for providing initial samples of **3**. Partial financial support by the Deutsche Forschungsgemeinschaft (OW) and postdoctoral fellowships by the Deutscher Akademischer Austauschdienst (MR) and the Deutsche Forschungsgemeinschaft (UZ) are gratefully acknowledged.

**ligands**                                   **complexes**

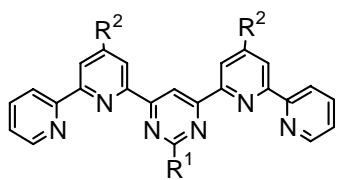

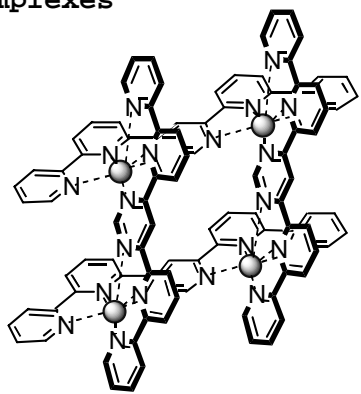

L^a  R^1 = Me    R^2 = H
L^b  R^1 = Br    R^2 = H
L^c  R^1 = H     R^2 = S^nPr

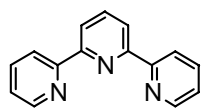

1   $[Co^{II}_4(L^a)_4](PF_6)_8$
2   $[Co^{II}_4(L^b)_4](PF_6)_8$
3   $[Co^{II}_4(L^c)_4](PF_6)_8$

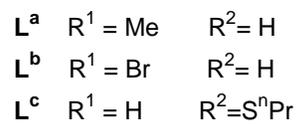

**tpy**

4   $[Co^{II}(tpy)_2](PF_6)_2$

**Figure 1.** Molecular structure of the Co(II)-[2 × 2] grid compounds **1-3** and of the mononuclear reference compound **4**.

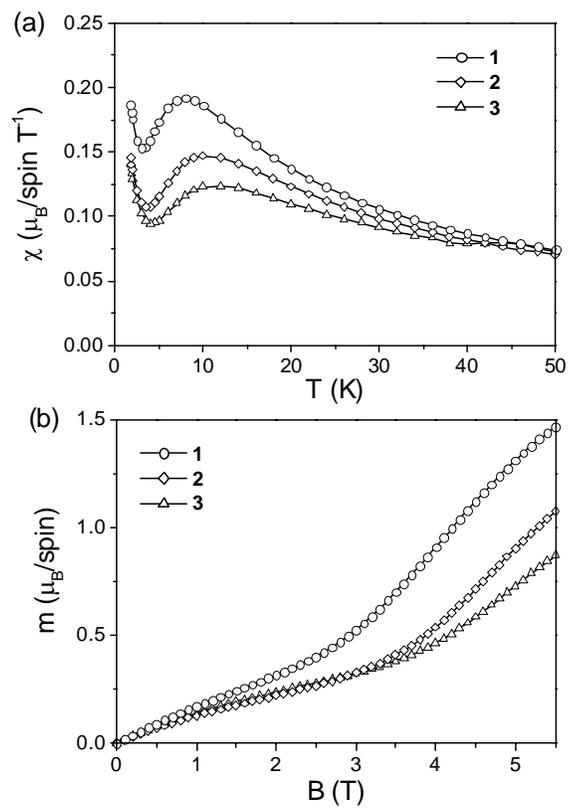

**Figure 2.** (a) Magnetic susceptibility vs. temperature and (b) magnetic moment vs. field at 1.9 K for powder samples of **1**, **2**, and **3**.

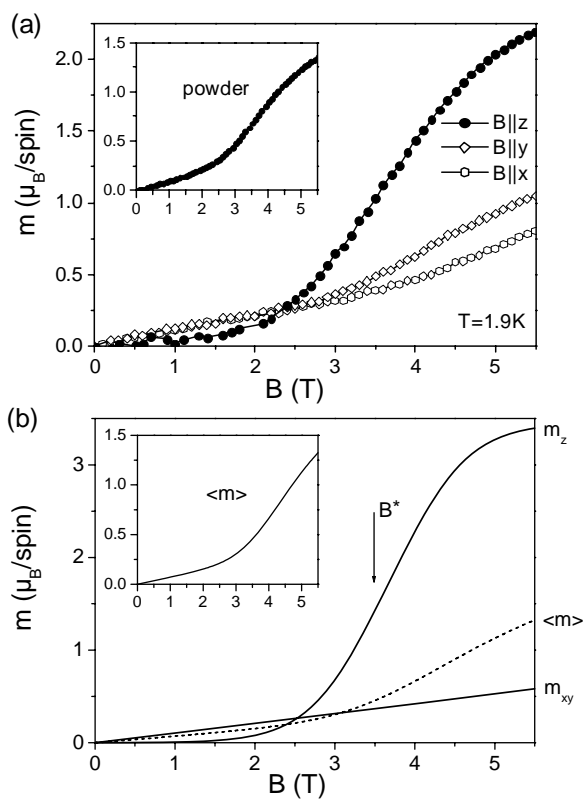

**Figure 3.** (a) Field dependence of the magnetic moment of a single crystal of **1** at 1.9 K for magnetic fields along the main axes. The inset shows the magnetic moment vs. field for a powder sample of **1**. (b) Magnetic moment vs. field as calculated with $\hat{H}_{2\times2}$ for J = -1.8 K, D = -20 K, and g = 2.3 at 1.9 K for fields in the z direction and the xy plane. The dashed line and the inset show the averaged magnetic moment $\langle m \rangle$ corresponding to powder samples.

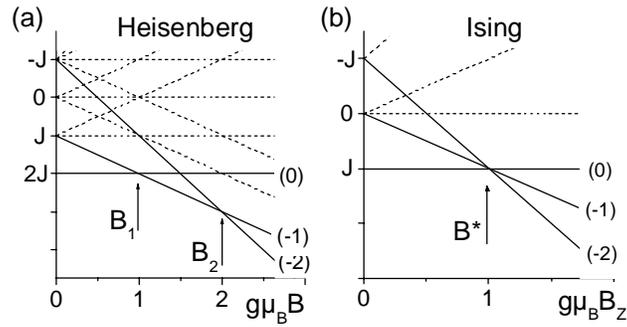

**Figure 4.** Energy spectrum for a [2 × 2] grid of spin-1/2 centers with antiferromagnetic (a) Heisenberg or (b) Ising interactions as function of magnetic field (along the z direction). Arrows mark the ground state level-crossings. Some states are classified by the total magnetic quantum number M.

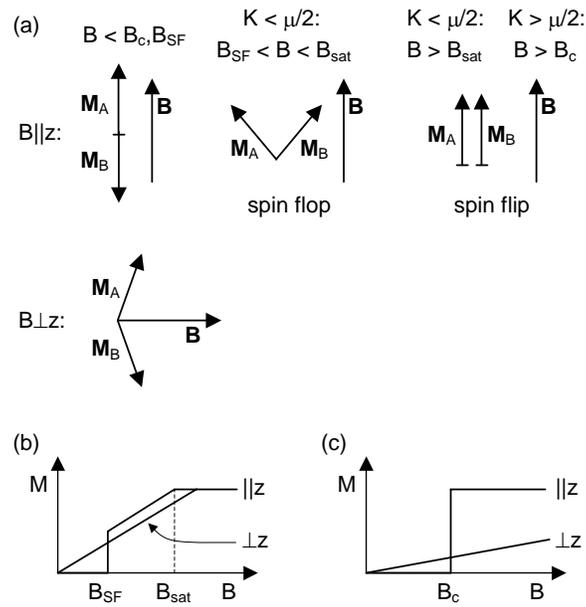

**Figure 5.** (a) Spin configurations of a classical antiferromagnet (for details see text). Panels (b) and (c) schematically show the magnetization curves for parallel and perpendicular fields of an antiferromagnet (with weak anisotropy) and a metamagnet, respectively.